\documentclass[11pt]{article}
\usepackage{apalike,latexsym,amssymb,bbm}

\def\titel{Fiber Bundle Gauge Theories and ``Field's Dilemma''}

\def\scmy{}         

\def\ben{\begin{enumerate}}
\def\een{\end{enumerate}}
\def\bi{\begin{itemize}}
\def\ei{\end{itemize}}
\def\bd{\begin{description}}
\def\ed{\end{description}}
\def\bq{\begin{quote}}
\def\eq{\end{quote}}
\def\bc{\begin{center}}
\def\ec{\end{center}}
\def\be{\begin{equation}}
\def\ee{\end{equation}}
\def\ba{\begin{array}}
\def\ea{\end{array}}

\def\setB{\mathbbm B}

\def\setE{\mathbbm E}
\def\setF{\mathbbm F}
\def\setH{\mathbbm H}

\def\setP{\mathbbm P}
\def\setR{\mathbbm R}
\def\setS{\mathbb S}
\def\setT{\mathbbm T}

\def\setV{\mathbbm V}

\def\setX{\mathbbm X}
\def\setY{\mathbbm Y}

\def\lie#1{\mathfrak #1}

\def\Mink{\setR^{1,3}}                
\def\Man{{\cal M}}                    
\def\Tan{{\cal T}}                    
\def\TM{{\cal TM}}                    
\def\DiffM{{\cal D\mbox{\em iff}(M)}} 
\def\AutP{{\cal A}ut(\setP)}          
\def\LM{{\cal LM}}                    

\def\Cur{{\cal C}}                    

\def\semisub{{\times \hspace{-1.1em} \subset}}

\begin{document}
\sloppy

\thispagestyle{empty}

\vspace*{-20mm}

\noindent {
Submitted to {\em Studies in History and Philosophy of Modern Physics}.}

\vspace{20mm}

\noindent{\Large\bf \titel}

\vspace{11mm}

\hfill\parbox{13cm}{
{\scmy Yair Guttmann}

\vspace{2mm}

{\small
Philosophy Department,
Stanford University,
Stanford, CA 94305, USA
}

\vspace{9mm}

{\scmy Holger Lyre}

\vspace{2mm}

{\small
Institut f\"ur Philosophie,
Ruhr-Universit\"at Bochum,             \\
D-44780 Bochum, Germany,
e-mail: holger.lyre@ruhr-uni-bochum.de
}

\vspace{9mm}

May 2000

\vspace{9mm}

\paragraph{Abstract.}
We propose a distinction between the physical and
the mathematical parts of gauge field theories.
The main problem we face is to uphold a strong and meaningful
criterion of what is physical.
We like to call it {\em ``Field's dilemma''},
referring to {\scmy Hartry Field}'s nominalist proposal
which we consider to be inadaequate.
The resolution to the dilemma, we believe,
is implicitly provided by the so-called fiber bundle formalism.
We shall demonstrate, in detail, that the bundle structure
underlying modern quantum and gravitational gauge field theories
allows for a genuine distinction between the physically significant
and the merely mathematical parts of these theories.
}\hspace*{5mm}


\newpage

\vspace*{9mm}


\bq
\renewcommand{\contentsname}{Contents}
\tableofcontents
\eq



\newpage

\section{The emergence of ``Field's dilemma''}

In his book {\em Science without Numbers} {\scmy Hartry Field} (1980)
raises an important ontological issue.
We believe in the existence of theoretical entities,
{\scmy Field} claims, because they are indispensable.
We need them in order to explain phenomena
which need to be explained.
However, our scientific theories in general and physical theories
in particular are formulated using, also, a mathematical vocabulary:
functions, transformations, integrals, derivatives and, of course,
real and complex numbers appear in the formulation of every
theory in physics. And, it seems that these mathematical entities,
too, are needed in order to construct explanations.
Does it not follow, then, that we should be committed to the
existence of these seemingly indispensable mathematical entities
for the same reason that makes us believe in the reality
of theoretical entities?
The answer, says {\scmy Field}, is that  the ``purely mathematical''
part of physics is not truly indispensable for explanatory purposes.
Let $T$ be the physical part of a theory and let $S$ be
its mathematical part. If $A$ is a physical fact which needs explanation,
says {\scmy Field},  and it can be explained in terms of $S+T$,
it can be explained in terms of $T$ alone.
The inclusion of $S$ is merely a matter of convention.
Hence, we do not have to worry about the ontology of its objects.

On this level of abstraction {\scmy Field}'s question strikes us as a very
important one and his general strategy for answering it seems promising.
Indeed, it seems that a significant advancement in our understanding of
physics will be achieved if we could identify and
characterize  the part of theoretical physics which stands for real
physical processes and distinguish  it  from  the merely mathematical and,
hence, the ``contentless'' part of physics to use {\scmy Field}'s own
terminology.
Our next step, then, is to outline {\scmy Field}'s specific strategy.
It is in this strategy that {\scmy Field} sees the merit of his account.

Let $T$ and $S$ be, respectively, the ``physical'' and
the ``mathematical'' parts of a theory.
Let $A$ be a ``nominalistically statable'' assertion $A$
($A$ is a statement whose quantifiers were restricted to range
only on non-mathematical objects).
{\scmy Field}'s strategy for demonstrating that $S$ is dispensable
is to prove that if $A$ is a semantic consequence of $S+T$
it is a consequence of $T$ alone.
Such a demonstration establishes that $S+T$ is a conservative extension
of $T$, namely that every model of $S+T$ is also a model of $T$ alone.
{\scmy Field} claims that if $T+S$ is proven to be a conservative
extension of $T$, $S$ is thereby shown to be ``contentless'';
it does not place any significant constraints on the models
of $T+S$ that could distinguish them from models of $T$ alone.
Indeed, writes {\scmy Field},
``{\em ... it would be extremely surprising if it were discovered that
standard mathematics implied that there are $10^6$
non-mathematical objects or that the Paris commune was defeated;
and were such discovery to be made all but unregenerate rationalists
would take this as showing that standard
mathematics needed revision. Good mathematics is conservative.}''
\cite[p.~13]{field80}.
Of course, says {\scmy Field}, we need to assume that standard mathematics
is consistent (otherwise $S$ would not have any non-trivial models).
This assumption, itself, cannot be proven conclusively.
However, we all believe that standard mathematics, being a very useful
theory, must be consistent.
Once this assumption is made, says {\scmy Field}, it is relatively easy to
prove that standard mathematics is conservative:
``{\em Indeed, ... the gap between the claim of consistency and
the full claim of conservativeness is, in the case of standard
mathematics, a very tiny one.}'' \cite[p.~13]{field80}.
This is certainly an interesting observation.
Regarding the issue of the truth of mathematics {\scmy Field} claims that
it is a mute point. Mathematics can be considered true if mathematical
entities are assumed to exist.
The main reason for believing in the existence of mathematical objects,
says {\scmy Field}, is that they are indispensable for explanatory purposes.
But since {\scmy Field} has supposedly shown that mathematical entities
are not needed for explaining physical facts
he regards the issue of mathematical truth as being simply irrelevant.

A critique of {\scmy Field}'s argument necessitates a full discussion
of his ``nominalistic reconstruction'' of various physical theories.
This undertaking, however, is not necessary in the present context; the
critique of {\scmy Field's} theory is not the main focus of this article.
What we want to present here is another way of answering
{\scmy Field}'s question, namely, how to divide physics into
a ``truly physical part'' and a ``merely mathematical'' part
and then demonstrate that the former should be regarded as representing
real and objective phenomena while the latter is to be regarded
as dispensable ``conceptual scaffolding''.
The details of our analysis, though, are significantly different.
Firstly, the sense in which physical theories are divided into
the physical and the mathematical parts is completely
different from the one {\scmy Field} employed.
{\scmy Field} uses logical tools to affect the decomposition;
we, on the other hand, shall use the mathematics of fiber bundles
to formulate our constructions.
Secondly, our characterization of what is physical is quite
different from {\scmy Field}'s.
For our purposes the physical part of the theory is the one
with experimental consequences. This is a much stricter criterion
than the one which underlines {\scmy Field}'s argument.
Because of our criterion we are led to consider the local gauge
covariance principle as the paradigm of what is physical.
We believe that this feature of our account brings us much closer
to the physicist's way of thinking.
Finally, the method of proving that various mathematical constructions
are dispensable vary significantly from {\scmy Field}'s.
First, we have nothing that is analogous to the proof of
conservative extension.
We believe that what makes the mathematical part of the theory different
from the physical part is not that we can do without it.
Instead, one should show that a different choice of mathematical
formulation would not have led to a theory with different experimental
consequences.

Ultimately, our account should be judged on its own merit;
it is not even absolutely clear that it is an alternative to
{\scmy Field}'s approach and not a complimentary set of observations.
We do, however, believe that {\scmy Field}'s account suffers from some
serious problems.
Consequently, we feel the need to ``save'' {\scmy Field's} general strategy
from the specific interpretation he assignes to it.
In this sense, and only in this sense, we are engaged in a critique
of {\scmy Field}'s account.

The problems in {\scmy Field}'s account to which we alluded are the following:
\bd
\item[]
(i) The formulations which {\scmy Field} offers to various physical theories
do not have much to be recommended for them, that is, apart from their
value for
the nominalist.
Although {\scmy Field} claims that his formulations yield ``reasonably
attractive'' theories we believe that physicists would find them quite
objectionable.

\item[]
(ii) The notion of semantic consequence which {\scmy Field} uses is
an abstract and non-constructive notion.
He does not show us how to actually derive any ``nominalistically statable''
assertion from $T$ alone. Hence, such assertions do not, stricly speaking,
receive an explanation from $T$ alone.

\item[]
(iii) {\scmy Field}'s idea of the physical is quite inflationary.
For instance, he regards spacetime points as physical entities.
In this respect his attempt to ``physicalize geometry''
is in stark contrast to the general-relativistically motivated
movement to ``geometrize physics''.

\item[]
(iv) More generally, the very logic of {\scmy Field}'s argument mandates
that he uses a very loose and weak notion of the physical so
that the resulting theory $T$ will be powerful enough to entail
the consequences of $T+S$.
This way of thinking is, again, in opposition to the line of thinking of the
physicist. As we remarked earlier, we believe that the term physical
assertion should be reserved
only for those aspects of physics which have empirical consequences.

\item[]
(v) Finally we do not believe, like {\scmy Field} does,
that we can do without mathematical concepts altogether
nor do we think that this is what we need to prove in order
to characterize the ontological difference between mathematical
and physical entities.
It might turn out that we cannot proceed without some mathematical
structures and notations.  What we need to demonstrate,
in order to account for the difference between the mathematical
and the physical, is that had we chosen another mathematical
``convention'' or construction, the experimentally testable part
of the theory would have remained essentially unaffected.
It is this type of invariance with respect to the choice
of mathematical description that allows us to conclude
that there is no need to interpret the ``merely mathematical''
entities of the theory as physically  real.
\ed

All these problems are quite troublesome.
There is, however, one issue that seems to us to deserve special attention.
In discussing what criterion to use for what is uniquely a physical
phenomenon many philosophers would tend to favor a relatively rigid
criterion; in particular, there is, clearly, much to recommend about
the idea that when we assert that a physical phenomenon occurs,
we must show that the assertion has empirical consequences.
On the other hand, in the interest of affecting a workable distinction
between the ``genuinely physical'' and the ``merely mathematical'',
some of the same philosophers might be susceptible to the pressure
of employing an inclusive criterion.
In this context even a loose characterization of what is physical,
if it yields a distinction between mathematics and physics,
seems preferable to no distinction at all.
This pressure is particularly strong for those who look for a proof
that the merely mathematical part of physics $S$ is dispensible.
By ``smuggling'' into the physical part $T$ large portion of $S+T$
one obtains a relatively strong $T$ and a relatively weak $S$,
making it easier to prove that $S+T$ is a conservative extension of $T$.
A similar motivation lurks behind much of what is attractive
in various forms of substantivalism.
When one regards space-time points as physical entities
one leaves very  little in the purely mathematical part
of physics. This residue seems to play a merely faciliatory role
akin to that played by the logical connectives;
therefore the inflationary criterion for the physical
leaves us with very little
temptation to attach to it any ontological significance.
The price, though, is in our opinion much too high.
Therefore, we shall try to find a way around it.

To sum up, then, we are faced with a double pressure.
On the one hand we would like to employ a relatively restrictive
criterion of what is physically significant;
but on the other we are pressured in the opposite direction
to affect a distinction between the mathematical and the physical parts.
This is what we propose to call {\bf\em ``Field's dilemma''}.
A large portion of this article is dedicated to show how to escape
this dilemma, that is, how to uphold a restrictive criterion
of what is physical while making it possible to distinguish beween
the mathematical part of physics and its truly physical parts.

\medskip

Let us now describe the contents of this paper.
Our main purpose is to present a new conception
of the ontology of gauge field theories (the most fundamental
field theoretic approach of physics today).
This conception, we believe, is implicit in the fiber bundle formalism.
The construction of fiber bundles makes it possible
to distinguish the ``merely mathematical'' aspects
of physical theories from their ``truly physical'' parts.
From this aspect stems the importance of fiber bundles
for the discussion of the ontology of physical theories, as we see it.
The gauge principle has a precise and uniform representation
in fiber bundle theoretic terms.
Therefore, a careful study of these structures will allow us
to better understand this principle and to distinguish between
gauge principle and ``merely mathematical'' covariance principles.

To those who are acquainted with the bundle formalism
there is no need to exalt its virtue at length.
It is enough to mention that, at least in principle,
every physical field theory can be given a specific formulation
using the construction of the appropriate fiber bundle.
Therefore, one may think of the fiber bundle formalism
as the {\em lingua universalis} of modern physics.
The basic ideas and definitions of bundle theory
have been present in the mathematical literature for well over fifty years.
These ideas are classified under the headings of differential geometry
and algebraic topology, two disciplines which gave rise to some of
the most important mathematical developments of the twentieth century.
However, the application of these ideas to physics
are much more recent. Only in the last two or three decades did
the power of these methods, as they apply to mathematical physics,
become evident. Mathematical physicists who used them managed to derive
impressive results which could not have been obtained in other ways.
Since then the study of fiber bundles has become part of the
standard education of young physicists. These developments lend
some urgency to the project of properly understanding the ideas
behind the fiber bundle formalism.
The investigation of these ideas and their applications to physics
will occupy the center part of this paper.


\newpage

\section{Fiber bundles}

In this chapter we try to give a brief and self-contained introduction
to the mathematics of fiber bundles. Our key motivation is the idea
of generalizing the direct product and its set theoretic background;
in the later section of this chapter we shall show the connection of
this presentation with more physics-oriented presentations such as
Drechsler and Mayer (1977) and Trautman (1984).


\subsection{The direct product and its generalization}
\label{generalization}

Of the many contributions of set theory to the foundations of mathematics
the set-theoretic definition of mathematical functions
which emerged early in the development of set theory
stands as a milestone.
Until the end of the 19th century mathematicians thought of functions
as rules of correspondence.
That is, if $x \in \setX$ is a member of the domain of $f$,
the expression f(x) was considered to be a procedure or a recipe
to obtain $y = f(x)$ from $x$.
As a consequence, in order to determine whether $y = f(x)$ or not,
one had to inquire into the meaning of the instruction underlying $f$.
Such inquiries, by definition, cannot be completely precise;
therefore, a need was felt for a more rigorous definition
of the concept of a mathematical function.

As the reader undoubtedly knows, in set theory the function
$f: \setX \to \setY$ is identified with its ``graph'' $G_f$, that is,
with the set of all ordered pairs\footnote{An ordered pair
can be identified with a simple set using, for example,
the convention $(x,y)= \{ x,\{y\} \}$.}
$(x,y)$ such that $f(x) = y$
(note that $G_f$ is a subset of the direct product $\setX \times \setY$).
Hence, the question of whether $y = f(x)$ can be
answered by simply checking that $(x,y) \in G_f$;
imprecise ``instructions'' and ``rules of correspondence''
are banished from the ontology of mathematics.
If $f$ is a well defined function, $G_f$ is a well defined set and the
question whether or not $(x,y) \in G_f$ has an unambiguous answer.

As a result of the following considerations, rather than
investigating $f$, we may investigate the relations between
$\setX,\setY$ and $\setX \times \setY$. More precisely, we investigate
\be
\label{dirprod}
\ba{rrcll}
      &   \setX  & \times & \setY    \\
\pi_x & \swarrow &        & \searrow & \pi_y \\
\setX &          &        &          & \setY
\ea
\ee
where $\pi_x(x,y) = x$ and $\pi_y(x,y) = y$.
$\pi_x$ and $\pi_y$ are called projections.
In view of the foundational significance of the direct product
construction it should be interesting for philosophers of mathematics
to find out how it can be generalized. Indeed, this is the motivating
idea behind the fiber bundle formalism. As we shall see shortly,
the generalization of the direct product can be viewed as yet
another conception of the nature of mathematical functions.

The fundamental idea of the fiber bundle construction is
to continue investigating $\setX$, $\setY$, $\setX \times \setY$
and $\pi_x$ but to give up $\pi_y$, the projection into $\setY$.
What could be a reason for giving up $\pi_y$?
The idea is that $\setY$ has an ``additional  layer of hidden structure''
which is ``indispensable'', that is, in order to determine the values
of $f$ this layer of structure has to be taken into account.
Therefore, an adequate representation of the situation should make
the dependence on this layer of structure explicit.
Let us think of the above mentioned layer of structure as a group $G$
which operates\footnote{We can think of the operation of $G$
as an equivalence relation $\sim$ which is defined on $\setY$.
If $y=gy'$ with $g \in G$, then $y \sim_g y'$.
In general, this equivalence relation ``decomposes'' $\setY$ into
equivalence classes, one for each of the elements which generate $G$.
Each of the equivalence classes is the $G$-orbit of its members.}
on $\setY$. We get
\be
\ba{rrcll}
      &   \setX  & \times & \setY    \\
\pi_x & \swarrow &        &          \\
\setX &          &        &          & G \times \setY \to \setY
\ea
\ee
instead of (\ref{dirprod}).
Now let $x \in \setX$ and let $\setY_x = \pi_x^{-1}(x)$
($\setY_x$ is called the fiber over $x$).
Let us now define a family of maps $\setY_x \to \setY$.
If $G$ is generated by a single element $g$ all the elements of $\setY$
are in one equivalence class and each of the elements of $\setX$
is assigned with $\setY_x$, an identical copy of $\setY$.
If $G$ is generated by more than one element for every generating
element of $G$ we shall have another map $\setY_x \to \setY$
(in other words, the assignment of $\setY_x$, the fiber over $x$,
is ``sensitive'' to the operation of $G$ on $\setY$).


\subsection{The definition of fiber bundles}

 This somewhat imprecise description will allow us to give the
reader a primary concept of fiber bundles and an idea of the way
they generalize the direct product.
A fiber bundle is a structure $\langle \setE, \setB, \pi, \setF, G \rangle $
which includes
\bd
\item[]
(i) The {\em bundle space} (or total space) $\setE$

\item[]
(ii) The {\em base space} $\setB$

\item[]
(iii) The {\em fiber space} $\setF$

\item[]
(iv) A mapping $\pi: \setE \to \setB$ called {\em projection}

\item[]
(v) A group $G$ called {\em structure group} with a left action on $\setF$

\ed
As we remarked $\setF_x = \pi^{-1}(x)$ is called the fiber over $x \in \setB$.

Let us turn now to the issue of the generalization of the direct
product. Is it, indeed, a genuine generalization?
If the point, where the definition of fiber bundles departs
from that of the garden variety of direct products,
is whenever the ``hidden structure'' over $\setY$,
which is represented by the action of the group $G$,
is taken into account,
why can't we simply define the quotient space\footnote{The construction
of $\setY/G$ can be described as a choice of the ``representative element''
from each of the equivalent classes of $\setY$ induced by $G$.}
$\setY/G$ and a function $f':\setX \to \setY/G$
which is to be investigated instead of $f$?
Note that if we want to claim that behind the fiber bundle formalism
there is a new conception of mathematical functions
there is some urgency to answer this question in a convincing way.
Is the fomalism we develop a genuinely new idea or is it merely
a hyphaluted mathematical notation?

 There are two reasons for believing that we are engaged in a
genuinely new body of ideas. First, as we shall see shortly,
not every fiber bundle is equivalent (in a precise mathematical
sense which we shall define soon) to a direct product.
Secondly, the ``hidden structure'',  which is coded by the action
of the group $G$, should not be ignored;
but we should not give up $\setY$ and concentrate only on $\setY/G$ either.
As we shall see later the invariance with respect to the action of $G$
often represents highly non-trivial physical facts.
Thus, we should be prepared to retain the ``superfluous structure''
of $\setY$ and think of $G$-covariance as a physical principle.
If we rid ourselves of the ``excess structure'' too quickly
we shall not be able to appreciate the difference between
this covariance with respect to $G$ and other more trivial
kinds of covariance without a physical meaning.
The full significance of the last remarks will become clearer
when we discuss some physical examples.


\subsection{Different types of fiber bundles}

\subsubsection{Sections and coordinate bundles}

For a direct product, i.e. a structure
$\langle \setX, \setY, \setX \times \setY, \pi_x, \pi_y \rangle$,
the following construction is always possible:
let $y_0 \in \setY$ and
\be
\setX_{y_0} = \left\{ (x,y) : \pi_y(x,y) = y_0 \right\}.
\ee
The set $\setX_{y_0}$ is always well defined.
$\setX_{y_0}$ is called a {\em global section} of $\setX \times \setY$.
If $\setX_y$ and $\setX_{y'}$ are global sections corresponding to
$y, y' \in \setY$, there is a natural isomorphism
$\varphi_{y,y'}: \setX_{y_0} \to \setX_{y_0}$, $(x,y) \to (x,y')$.
There is also a natural isomorphism $\varphi_y: \setX \to \setX_y$
between $\setX$ and $\setX_y$, $x \to (x,y)$.
Hence we may regard the various $\setX_y$ as
``copies'' of $\setX$ attached to the various $y \in \setY$.
In case we give up $\pi_y$, we give up the possibility of always
having global sections at our disposal.
Instead we make a weaker requirement, namely, that for every
$x \in \setX$ there is a neighborhood $U_x$ such that in the
``sub-bundle'' $U_x \times \setY$ we can construct {\em local sections}.
This requirement is called a {\em local trivialization}.
As we shall see shortly the local trivialization requirement is,
indeed, a weaker condition which
does not always guarantee the existence of global sections.

Let us describe with some detail how to construct fiber bundles
which satisfy the local trivialization condition.
In such cases, each point of the base space is contained
in an open neighborhood $U$ such that the portion of the bundle
``above'' $U$ is isomorphic to a direct product.
Note that in order to define such a fiber bundle
we must confine our attention to those cases where the notion of
an open neighborhood is well defined, that is,
to differentiable manifolds or, at least, to topological spaces.

Recall that a {\em chart} on a topological space $\Man$ is
a homeomorphism $f: U \to V$ of $U \subset M$, an open set,
onto an open subset $V \subset \setR^n$
($n$ is called the dimension of $f$).
An {\em atlas} is a collection of charts whose domains cover $\Man$.
$\Man$ is an $n$-dimensional manifold if it admits an $n$-dimensional atlas.
Consider two charts $f:U \to V$ and $f':U' \to V'$ on an $n$-dimensional
manifold $\Man$. We say that $f$ and $f'$ are {\em $C^\infty$ compatible}
if the composit maps
$f \circ f^{\prime \ -1} :  f(U \cap U') \to (U \cap U')$
and
$f' \circ f^{-1} :  (U \cap U') \to (U \cap U')$
are of class $C^\infty$ (that is, they have well defined partial derivatives
of all orders.) An atlas is of class $C^\infty$ if all its charts are
pairwise $C^\infty$ compatible. The construction of a $C^\infty$ atlas
shows how to introduce a coordinate system into a topological space.
We use local charts to ``import'' into $\Man$ the coordinate system
of $\setR^n$; these coordinates enable us to assign the objects
of $\Man$ a location.
The fact that the charts are $C^\infty$ compatible guarantees a
``smooth'' transition from one coordinate patch to another.

Using an analogous procedure we can show how to construct
a fiber bundle from local patches. Let $\setB$ be the base manifold
and $\{U_n\}$ a collection of subsets of $\setB$ which cover $\setB$.
Let $\setE$ be a fiber bundle with $\setB$ as the base space,
$\setF$ the ``typical fiber'' and $\pi: \setE \to \setB$ a projection.
Observe that for all $n$, $\pi^{-1}(U_n)$ is a subset of $\setE$
(clearly $\bigcup_n \pi^{-1}(U_n)$ covers $\setE$).
If we restrict our attention to locally trivial bundles
we can assume that there is a diffeomorphism
$\phi_n: \pi^{-1}(U_n) \to U_n \times \setF_n$  where $\pi^{-1}(U_n)$
is a subset of $\setE$ and $\setF_n$ is the fiber associated with
some $b \in U_n$.
Now consider two patches $U_n$ and $U_m$ with a non empty intersection.
Let $\phi_{nm}$ be the restriction of $\phi_n$
to the $\pi^{-1}(U_n \cap U_m)$, that is
$\phi_{nm}: \pi^{-1}(U_n \cap U_m) \to (U_n \cap U_m) \times \setF_n$.
Similarly let $\phi_m: \pi^{-1}(U_m) \to U_m \times \setF_m$ and let
$\phi_{mn}: \pi^{-1}(U_m \cap U_n) \to (U_m \cap U_n) \times \setF_m$
be the restriction of $\phi_m$ to $\pi^{-1}(U_n \cap U_m)$.
Let us now define the composit diffeomorphism $\rho_{nm}$
\be
\rho_{mn} = \phi_{nm} \circ \phi^{-1}_{mn}:(U_n \cap U_m) \times \setF_n
            \to (U_m \cap U_n) \times \setF_m ,
\ee
which is called the {\em transition function}
between the bundle charts $\phi_{nm}$ and $\phi_{mn}$.

Let us now consider three open patches $U_n$, $U_m$ and $U_l$
and restrict our attention to the intersection $U_n \cap U_m \cap U_l$.
We need to know how to switch from one location to another.
To this end we will define the transition functions $\rho_{nm}$,
$\rho_{ml}$ and $\rho_{nl}$ which correspond to the transitions
$U_n \to U_m$, $U_m \to U_l$ and $U_n \to U_l$ respectively.
The transition from one coordinate patch to another should not depend
on the path. Hence we require that
\be
\rho_{nm} \circ \rho_{ml} = \rho_{nl} .
\ee
This condition which is called the {\em cocycle condition},
makes it possible to define, for each transition,
a unique inverse transition. As we shall see shortly
the cocycle condition allows us to code the various
transition functions as elements of a structure group $G$.
More precisely the condition makes it possible to associate
with each $g \in G$ the inverse element $g^{-1}$
such that $g g^{-1} = id$.

  To sum up, suppose that we are given a manifold $\setB$
with an open cover $\{U_n\}$ and we assign to each $b \in U_n$
the fiber $\setF_n$. Suppose further that for each $n,m$ we can
define the transition function $\rho_{nm}$ and that
the collection of such function satisfies the cocycle condition.
In such a case we can define a fiber bundle $\setE$
by patching together the sets $U_n \times \setF_n$
by means of the transition functions $\rho_{nm}$.
It can be shown that $\setE$, thus obtained,
is a metrizable separable and locally compact space.
In fact, $\setE$ is, itself, a differentiable manifold;
the domains of the charts of $\setE$ are $U_n \times \setF$.
$\pi: \setE \to \setB$ is defined on each chart as a projection
on the first element of the product.


\subsubsection{Principal bundles and associated vector bundles}

Our next step is to construct principal fiber bundles
where the action of the structure group is explicit.
We shall do so by requiring that each of the fibers can be
identified with the group $G$.
In more pedestrian cases one must distinguish between $G$,
taken as a set of operations which can be composed with one another
and the domain of objects on which the members of $G$ operate.
However, in the present case we are considering mathematical objects
with a dual character.
On the one hand we are dealing with a topological space,
on the other hand each $g \in G$ is an operation.
The composition  $G \times G \to G$ is defined as
$(g_1,g_2) \to g_1 g_2^{-1}$.
It is required to be continuous with respect to
the product topology on $G \times G$.
If $G$ has, in addition, the structure of a differentiable manifold,
then it is natural to require that the composition is smooth.
When this requirement is satisfied G is called a {\em {\scmy Lie} group}.
Another requirement on $G$ is that it acts freely;
that means that it does not have non-trivial fixed points.

A {\em principal bundle $\setP$}
is a structure $\langle \setP, G, \setB, \pi \rangle$
where $\setP$ is a manifold on which the group $G$ acts freely.
The projection $\pi: \setP \to \setB$ is a $C^\infty$ function
from $\setP$ onto $\setB$. $\setP$ is assumed to fulfil
the local trivialization condition: for every $b \in \setB$
there is an open set $U \subset \setB$ and an isomorphism
$f: U \times G \to \pi^{-1}(U)$ such that for every
$b \in U$ and $g,g' \in G$ we get
$\pi(f(b,g)) = b$ and $f(b,g g') = f(b,g) g'$.
Note that a tight connection exists between the projection $\pi$
and the orbits of $G$.
All the elements of $\setP$ which are projected onto the same
$b \in U$ are transformed into one another by the elements of $G$.
In other words, the fibers of $\setP$ are the orbits of $G$ and,
at the same time, the set of elements which are projected
onto the same $b \in U$.
This observation motivates calling the action of the group
``vertical'' and the base manifold ``horizontal''.
We shall explain this choice of terminology further shortly.
 
As we remarked earlier the requirement of local trivialization is not
sufficient to guarantee that the fiber bundle is mathematically trivial.
To clarify this point let us state necessary and sufficient conditions
for a fiber bundle to be trivial. In the case of principal
fiber bundles these conditions are rather intuitive.
Recall that a section of $\setP$ is a diffeomorphism
$s : \setB \to \setP$ such that $\pi s = id$.
Now, $\setP$ is isomorphic to the trivial bundle
$\langle \setB \times G, G, \setB, \pi \rangle$
if and only if it admits a global section.
Indeed, consider the map $f : \setP \to \setB \times G$
defined by $f^{-1}(b,g) = s(b)g$ for all $b \in \setB$, $g \in G$.
$f$ is a $C^\infty$ bijection and, hence, an isomorphism.
Note that, as far as principal bundles are concerned,
the base manifold can be considered as the quotient $\setP/G$.
This point of view is equivalent to the construction using coordinate
patches. Historically, it was not immediately evident that that
the definition of principal fiber bundles in terms of coordinate patches
is equivalent to defining them in terms of a quotient space.
 
To facillitate our understanding of the notion of mathematically
trival fiber bundle let us take two examples.
The base space, in both cases, is a circle $\setS^1$
with two coordinate patches $U$ and $U'$
(in order to cover a circle we need, at least, two patches).
The fiber is, in both examples, the unit interval $[0,1]$.
In the first example the structure group $G$ has two generators
$e$ and $r$. $e$ is an infintesimal translation
and $r$ is a rotation of the unit interval around its center.
Because of the action of $r$ after a full circle the unit interval
will be rotated by 180 degrees. Thus, the zero point will be identified
with the other pole 1 and, vice versa, 1 will be identified with 0.
As a consequence the construction will result in a {\scmy M\"obius} strip,
a non-trivial fiber bundle which does not admit a global section.
In the second case the structure group $G'$ has only one generator:
the infintesimal shift $e$.
The result, in this case, is a cylindrical fiber bundle;
the cylinder is a trivial fiber bundle which admits sections
(however, as we shall discuss later in greater detail, bundles
which are trivial mathematically might be far from trivial
from a physical point of view).

Once we have defined a principal fiber bundle we can make an extra step
and construct fiber bundles which are associated with the principal bundle.
The general idea is to represent the structure group in another structure
(which is intended to be used as a ``typical fiber''), and assign copies
of this structure to the points in the base space of the principal bundle.
Because the main focus of this paper is not purely mathematical, 
rather then defining an associated fiber bundle in the most general way
we shall define the vector bundle associated with a principal bundle.
One example of an associated vector bundle is the tangent bundle.
Vector bundles appear frequently in physical applications.
Let $\setV$ be a vector space of dimension $n$,
then, a representation of $G$ in $V$ is a mapping
\be
    \rho : G \times \setV \to \setV ,
\ee
$(g,v) \to v'$ defines the action of $G$ on $\setV$.
Let $\setP \times G_\setV = (\setP \times \setV)/G$.
Now we can define the structure
$\langle \setP \times G_\setV, G, \setB, \pi \rangle$ to be the
{\em vector bundle $\setE$ associated with the principal bundle $\setP$}.
When we assign to each point $p \in \Man$ its tangent space $\setT_p$,
the union $\bigcup \setT_p$ is the tangent bundle $\TM$
over the base manifold $\Man$.
We shall discuss the tangent bundle further shortly.


\subsection{Connections on a principal fiber bundle}

Connections are needed in order to formulate a ``law''
which determines how various objects are transported
from one point of a manifold to another.
Even when we know how objects evolve along the base manifold
(that is, if we know the spatiotemporal coordinates of the evolution),
we still need to determine the evolution along the bundle manifold.
Hence, we need a method or a principle telling us how to ``lift''
a curve from the base manifold to the fiber bundle.
Such a principle should enable us to define the notion
of parallel transport. We shall discuss how to formulate
these notions for a vector bundle associated with a principal bundle.
This formulation is not the most general one but it contains
most of the ideas one encounters in the physics of fiber bundles.

Now, we shall think of $\setP$ as the bundle of linear frames $\LM$.
Let $u \in \setP$ and $p \in \Man$ such that $\pi(u) = p$.
We wish to transport a vector from its origin at $p$
to neighboring points along vectors emenating from $p$.
Let $v \in \setT_p\Man$ be such a vector
($\setT_p\Man$ is the collection of vectors tangent to $p \in \Man$).
We expect the law of transport to fulfil the following conditions:
\bi
\item[(i)]
It should depend smoothly on $p$ (that is, if $p'$ is infintesimally
near $p$ the evolution from $p'$ can be expressed as an infintesimal
variation on the evolution from $p$).

\item[(ii)]
The law should allow us to define the parallel transport along
any vector $v \in \Tan_p\Man$.

\item[(iii)]
If $v$ is transported from $u$ to $u'$ then, if $g \in G$, $ug$
should be transported to $u'g$.
\ei

Formally, in order to establish a connection, a vector space
$\setH_u\setP \subset \setT_u\setP$ must be assigned to every
$u \in \setP$ ($\setT_u\setP$ is the vector space of the tangents
at $u$, that is, $\setT_u\setP \in \setT\setP$).
The conditions (i)-(iii)  can now be formulated more precisely.
\bi
\item[(i)*]
The assignment must be smooth.

\item[(ii)*]
Let us define $\Theta_u\pi: \setH_u\setP \to \setT_{\pi(u)}\Man$.
We require that $\Theta_u\pi$ is an isomorphism for every $u \in \setP$.

\item[(iii)*]
The assignment should be invariant with respect to the action of $G$:
if $g \in G$ and $\Phi_g(\setH_u\setP)$ is the result of the action of $g$
on the members of $\setH_u\setP$ then $\Theta_u\Phi_g(\setH_u\setP) = \setH_{ug}\setP$.
\ei
Note that the isomorphism $\Theta_u\pi$ allows us to ``lift''
any vector $v \in \setT_p\Man$ to a unique vector in $\setT\setP$.
More precisely, if $\pi(u) = p$ and $v \in \setT_p\Man$
we may define the lift $\lambda_u v \in \setH_u\setP$ by setting
$\lambda_u v = (\Theta_u\pi)^{-1}(v)$. We may define
\be
\setV_u\setP = \{ v \in \setT_u\setP \ | \ \setT_u\setP (v) = 0 \}
\ee
as the space of all vectors tangent to the fiber of $\setP$ through $u$.
This fact motivates the idea that $\setV_u\setP$ is the ``vertical'' part
of $\setT_u\setP$. Hence, the direct sum decomposition
\be
\label{decomp}
   \setT_u\setP = \setV_u\setP \oplus \setH_u\setP
\ee
together with the isomorphism between $\setH_u\setP$ and $\setT_p\Man$
justifies calling $\setH_u\setP$ the ``horizontal'' part of $\setT_u\setP$.
Now, a connection on $\setP$ allows for a unique separation
of the vertical and the horizontal part of $\setT_u\setP$ according
to (\ref{decomp}). Let $\lie{g}$ be the {\scmy Lie} algebra of $G$,
then the connection is defined as a $\lie{g}$-valued one-form projecting
$\setT_u\setP$ to $\setV_u\setP \cong \lie{g}$.

Finally, we can define the horizontal lift of a parametrized curve
$\Cur: [0,1] \to \Man$. Let $\Cur(0) \in \Man$ and $u \in \setP$
with $\pi(u) = \Cur(0)$.
A horizontal lift $\bar\Cur: [0,1] \to \setP$ of $\Cur$ satisfies:
(i)   $\bar\Cur(0) = u$,
(ii)  $\pi(\bar\Cur) = \Cur$,
(iii) the tangent to $\bar\Cur$ at $t \in [0,1]$
      is a member of $\setH_{\bar\Cur(t)}$.
As we remarked earlier, the isomorphism between $\setH_{\bar\Cur(t)}$ and
$\setT_{\Cur(t)}\Man$ justifies calling $\bar\Cur$ a horizontal lift of $\Cur$.
One may say that $\bar\Cur$ is obtained by a parallel transport of $u$
along $\Cur$. Generally, the fact that $\Cur(0) = \Cur(1)$, that is,
that $\Cur$ is a closed loop does not guarantee that the horizontal
lift $\bar\Cur$ is a closed loop as well.
It might be the case that $\bar\Cur$ is not a closed loop and
$\bar\Cur(0) \ne \bar\Cur(1)$.
In general, if $\Cur(0) = \Cur(1)$ then $\bar\Cur(1) = \bar\Cur(0) \ g$,
where $g \in G$. The set of all $g \in G$ which can be obtained
in this manner forms a subgroup of $G$ called the {\em holonomy group}
of the connection at point $\Cur(0)$.

\bigskip

We have now introduced the essential notions of fiber bundle
mathematics and shall proceed to their applications
in gauge field physics.


\newpage

\section{Gauge field theories}

The previous section of this paper was devoted to the general
mathematical concept of fiber bundles.
In this section we will discuss the advantages of using
fiber bundles in physics.
Our demonstration is set in the realm of modern gauge field theories.
It is in this area of physics,
perhaps the epitome of physics today,
where our key assertion must be proven.
We need to show that the significant quantities
of gauge field theories ``live in bundle spaces''
rather than the spacetime base manifold.


\subsection{General conceptual and terminological issues}

Let us start with a list of the main concepts occuring
in gauge field theories. It will be useful to clarify
the terminology needed for later discussions.\footnote{We
also address some terminological issues which were first
presented in Lyre (1999).}

\bd
\item[(i) Covariance and invariance.]
The notions of covariance and invariance, which are crucial
to our considerations, have been given different definitions
and interpretations throughout the literature
(in physics textbooks they are often used interchangeably).
We shall use the following terminology:
covariance means {\em form invariance of the equations
of the theory with respect to some group of transformations},
the so-called {\em covariance group.}
In gauge field theories, we better speak
of {\em gauge covariance} instead of gauge invariance.
We shall therefore make a distinction between covariance and a looser
notion of a {\em group invariance of certain objects} of a given theory,
which does not involve the form of the equations.

\item[(ii) {\scmy Noether}'s theorem.]
In order to understand gauge theories we must study
the connection between global symmetries and conserved quantities.
Such a connection is established by {\scmy Emmy Noether}'s first theorem:
\bq
{\em
{\em {\scmy Noether}'s theorem:}
Let $\phi_i(x)$ be a field variable
($i$ is the index of the field components),
then the covariance of the action functional
$S[\phi]=\int {\cal L} [ \phi_i(x), \partial_\mu \phi_i(x) ] \, d^4x$
under a $k$-dimensional {\scmy Lie} group implies
the existence of $k$ conserved currents.
}
\eq

\item[(iii) Gauge postulate.]
The defining feature of gauge field theories is that they
couple matter fields and interaction fields dynamically.
The beautiful idea of the gauge approach is to start with a free theory,
and then ``derive'' the structure of the coupling from the
assumption that the following postulate is satisfied:
\bq
{\em
{\em Gauge postulate:}
The Lagrangian of a free matter field $\phi_i(x)$
should remain invariant under local gauge transformations
$\phi_i(x) \to \phi'_i(x) = \phi'_i [\phi_i(x), \alpha_s(x)]$.
}
\eq

\item[(iv) Gauge principle.]
The idea of gauging rests on postulating local gauge covariance
instead of the corresponding global {\scmy Noether} covariance.
This idea\footnote{The general idea in a compact manner was first
introduced by {\scmy Hermann Weyl} in his seminal paper from 1929;
cf. O'Raifeartaigh (1995) for historical remarks.}
is captured in the {\em principle of local covariance}:
\bq
{\em
{\em Gauge principle:}
The coupling of the {\scmy Noether} current corresponding
to the global gauge transformations of the Lagrangian of free matter
fields can be introduced as the replacement of the usual derivative
by the covariant derivative $\partial_\mu \to D_\mu$
which corresponds to local gauge transformations.
}
\eq

\item[(v) Gauge transformations.]
We must distinguish between
{\em global} gauge transformations (GTG) and
{\em local}  gauge transformations (GTL).
Regarding GTL, we must further distinguish two
subcategories\footnote{Unfortunately, the terminology in the
literature is not uniform. Some textbook authors already call
global and local gauge transformations transformations of the
first and second kind (compare Ryder (1985, chap.~3.3), for instance).
But then they lack to distinguish between matter field and
gauge potential transformations. Therefore, we prefer the above
terminology originally introduced by Pauli (1941).}:
\bi
\item[a)]
   matter field    transformations, also called
   ``gauge transformations of the first  kind'' (GTL1), and
\item[b)]
   gauge potential transformations, also called
   ``gauge transformations of the second kind'' (GTL2).
\ei
In quantum gauge field theories, the matter fields of transformations
GTL1 describe the fundamental fermionic particles of the standard model,
such as leptons and quarks\footnote{A more precise term would be
``energy-matter fields'', since there may exist fundamental
particle fields with mass zero such as, perhaps, neutrinos.
Most precisely, since there are massive gauge particles, such as W bosons,
we should speak about ``energy-matter fields which are no gauge fields''.
The reader may read our term ``matter field'' as a shortform.},
whereas in gravitational gauge theories the ``matter fields''
are tangent vector fields associated with {\em reference frames}
(that is, a material system representing an observer,
measuring rods, and clocks).

\item[(vi) Bundle structure of gauge field theories.]
Modern gauge field theories are geometrically characterized
as principal fiber bundles $\setP$, where the gauge fields ``live'',
and their associated vector bundles $\setE$,
in which the matter fields ``live''.
Thus, in bundle theoretic terminology, the distinction between
GTL1 and GTL2 (and their respective fields) reflects
the distinction between $\setE$ and $\setP$.
Gauge potentials are connections on $\setP$, local sections
of $\setE$ represent matter fields. The gauge group $G$ is the structure
group of the bundle and its generators represent the gauge bosons.
The group of local gauge transformations $\cal G$ may be regarded
as the automorphism group of $\setP$.
Finally, the bundle curvature is to be considered as the
interaction field strength.
We may, hence, compose a brief dictionary for the bundle terminology
of gauge theories as shown in the table on page \pageref{our_table}.

\begin{table}[t]
\label{our_table}
\fbox{\parbox{158mm}{
\bc
\begin{tabular}{l|l}
Fiber bundle framework            & Gauge field theory \\
\hline
principal bundle $\setP$          & geometric arena of gauge potentials \\
connections on $\setP$            & gauge potentials \\
associated vector bundle $\setE$  & geometric arena of matter fields \\
local sections of $\setE$         & matter fields   \\
structure group $G$               & gauge group  \\
generators of $G$                 & gauge bosons \\
automorphisms of $\setP$          & local gauge transformations GTL \\
vertical automorphisms of $\setP$ & pure  gauge transformations \\
covariant derivative              & dynamical coupling \\
curvature                         & interaction field strength
\end{tabular}
\ec
}}
\vspace*{-1mm}
\bc
{\sf Table: Comparison of bundle theoretic and gauge theoretic terminology}
\ec
\end{table}

\item[(vii) Gauge freedom.]
It is common practice to call classical electrodynamics,
i.e. {\scmy Maxwell}'s theory, already a ``gauge theory''.
The reason is that the {\scmy Maxwell} equations
are covariant under specific GTL2.
But this is certainly a misleading terminology; we shall rather
refer to it as a {\em gauge freedom} of the theory,
whereas only the {\scmy Dirac-Maxwell} theory,
or quantum electrodynamics (\ref{L_QED}), respectively,
combines the matter field and the gauge potential and, hence,
should be considered a true gauge field theory.

\item[(viii) Gauge field theory.]
The characteristic feature of any gauge field theory
is that it describes the coupling of a pure matter field theory
to a pure interaction ``field'' (or, rather, gauge potential) theory.
The structure of the coupling may be derived from the gauge principle
by imposing GTL covariance.\footnote{\label{gep} Recently,
the physical content of the gauge principle has been questioned
by several authors \cite{teller97,redhead98,brown99}.
In fact, the idea of imposing local gauge covariance leads,
strictly speaking, to just a special kind of bundle space
coordinate transformation. At this point, the occurrence
of a gauge connection has no physical significance.
Therefore, it can be argued that the gauge principle is not sufficient
to ``derive'' the coupling of matter and gauge fields and that a further
physical assumption is needed. As one of us has pointed out,
this assumption may perhaps be formulated in terms of a gauge theoretic
generalization of the equivalence principle \cite{lyre2000b}.}
The interaction potential, then, couples
to the matter field {\scmy Noether} current which is constructed
from the corresponding GTG covariance.
Any gauge theory represents the geometry of a principal fiber bundle
(and the associated vector bundle).
The gauge group is given by the structure group of the bundle.
\bq
{\em
{\em Gauge field theory} (for short: gauge theory):
A theory which describes the coupling of a matter field and
an interaction field. It is based on a gauge principle and
uniquely determined by its underlying principal bundle structure.
}
\eq

\ed


\subsection{Quantum gauge field theories}

At least three of the four known fundamental interactions undoubtedly
fit into the gauge theoretic framework. They are, moreover, suitably
formulated as quantum field theories. For our purposes, however,
the quantum field theoretic aspect does not play any special role.
The argument for this is that the structure of the Lagrangian
of a certain quantum field theory, quantum electrodynamics (QED)
for instance, is the same in the quantum field theoretic case
as in the case of its classical field theoretic counterpart.
Also, the bundle structure, which is our main concern here,
remains the same in both cases.
Thus, for our purposes QED is already captured in the {\scmy Dirac-Maxwell}
theory.\footnote{We should mention, however, that in the quantization
process of gauge theories certain technical problems arise.
This is because the considered interaction theory includes a gauge freedom
(thus, these problems refer primarily to GTL2 instead of GTL1).
But again, since these questions are not related to the bundle structure
of the theory we shall not be concerned with them.}
In the following two subsections we first derive the
{\scmy Dirac-Maxwell} theory from the gauge principle
in order to demonstrate its practical realization.
Next, we shall describe the extension of the gauge approach
to non-abelian gauge theories,
the so-called {\em {\scmy Yang-Mills} theories}.


\subsubsection{Dirac-Maxwell theory}

As a paradigm we consider the free {\scmy Dirac} field $\psi(x)$
with the {\scmy Lagrange} density
\be\label{L_D}
{\cal L}_D =
\bar\psi(x) \, \left(i \gamma^\mu \partial_\mu - m \right) \, \psi(x) .
\ee
The free {\scmy Dirac} Lagrangian is covariant under
global gauge transformations
\be\label{globaltrafo}
\psi(x) \to \psi'(x) = e^{iq \alpha} \psi(x) , \qquad
\bar\psi(x) \to \bar\psi'(x) = e^{-iq \alpha} \bar\psi(x)
\ee
with some arbitrary constant phase parameter $\alpha$ and charge $q$.
The {\scmy Noether} current corresponding to the transformations
(\ref{globaltrafo}) is given by
\be\label{noethercurrent}
\jmath^\mu = - q \, \bar\psi(x) \gamma^\mu \psi(x) .
\ee
It satisfies the continuity equation,
\be
\partial_\mu \jmath^\mu = 0 ,
\ee
which expresses the conservation of charge.
Following the gauge principle we replace the GTG in (\ref{globaltrafo})
by their corresponding GTL1
\be\label{phasetrafo}
\psi(x)     \to     \psi'(x) = e^{ iq \alpha(x)}     \psi(x) , \qquad
\bar\psi(x) \to \bar\psi'(x) = e^{-iq \alpha(x)} \bar\psi(x)
\ee
with a local, i.e. spacetime dependent phase function $\alpha(x)$.
This leads to the coupling to an interaction potential
\be
\label{coupling}
A'_\mu(x) = - \partial_\mu \alpha(x),
\ee
which itself satisfies GTL2
\be
\label{pottrafo}
A_\mu(x) \to A'_\mu(x) = A_\mu(x) - \partial_\mu \alpha(x).
\ee
Applying (\ref{phasetrafo}), (\ref{coupling}) and (\ref{pottrafo})
to (\ref{L_D}) yields to the covariant {\scmy Dirac} Lagrangian
\be
\label{L_D'}
{\cal L}'_D = {\cal L}_D + {\cal L}_{int}
\ee
with the coupling part
\be
{\cal L}_{int} = - \jmath_\mu(x) A^\mu(x) .
\ee
In accordance with the gauge principle,
this can directly be seen from the introduction of a
{\em covariant derivative}
\be
\label{cov_der}
\partial_\mu \to D_\mu = \partial_\mu - iq A_\mu(x) .
\ee
Thus, applying (\ref{cov_der}) to (\ref{L_D}) leads, again,
to (\ref{L_D'}).

The gauge theoretic framework necessitates an {\em active}
interpretation of the local gauge transformations GTL.
Note that ``active'' has the manifest meaning of
``physically significant''.
This should not be confused with the distinction between point and
coordinate transformations, which is also sometimes referred to
as a distinction between active and passive transformations.
The active character of GTL results from their changing
the physical interaction-free situation (\ref{L_D})
into an interaction coupling (\ref{L_D'}).
We will return to this issue in section \ref{resolve}.

Now, from the bundle theoretic point of view the vector field $A_\mu$
represents the components of the connection one-form of a $U(1)$
principal bundle. From (\ref{pottrafo}) we may as well construct
the corresponding curvature tensor
\be
\label{fmunu}
F_{\mu\nu}(x) = \partial_\mu A_\nu(x) - \partial_\nu A_\mu(x) .
\ee
Interpreting $A_\mu$ and $F_{\mu\nu}$ as potential and field strength
of the electromagnetic interaction, we are motivated\footnote{Of course,
this ``interpretation'' is not {\em enforced} by the gauge principle;
compare footnote \ref{gep}.}
to complete the Lagrangian (\ref{L_D'}) to the full QED Lagrangian
\be
\label{L_QED}
{\cal L}_{QED} = {\cal L}_D + {\cal L}_{int} + {\cal L}_{M}
\ee
with the free {\scmy Maxwell} field part
\be
\label{L_M}
{\cal L}_{M} = - \frac{1}{4} F_{\mu\nu}(x) F^{\mu\nu}(x) .
\ee

The gauge theoretic feature of a dynamical coupling of two field
theories into one combined framework is reflected in the existence of
two sorts of equations: equations of motion for the matter fields
(such as the {\scmy Dirac} equation which stems from ${\cal L}_D$
or, in gauge covariant manner, from ${\cal L}_D + {\cal L}_{int}$)
as well as interaction field equations
(such as the {\scmy Maxwell} equations which stem from ${\cal L}_M$
or, in inhomogeneous form, from ${\cal L}_M + {\cal L}_{int}$).
In the combined framework we call $A_\mu$ the {\em gauge potential}
and $F_{\mu\nu}$ the {\em gauge field strength}.
Quantum gauge field theories are usually formulated on a flat spacetime
manifold, i.e. {\scmy Minkowski} spacetime $\Mink$. Therefore,
the principal fiber bundle structure of QED is $\setP( \Mink,U(1) )$,
indicating that QED is a $U(1)$ gauge theory.


\subsubsection{Yang-Mills theories}
\label{ymt}

The gauge approach can be extended to non-abelian unitary
gauge groups.\footnote{This was first done by {\scmy C. N. Yang}
and {\scmy R. L. Mills} for the strong isospin $SU(2)_F$ \cite{yang+mills54};
cf. Mills (1989), O'Raifeartaigh (1995) for historical remarks.}
In the standard model case we use $SU(2)_L$ of weak isospin,
(unified in the elektroweak model with hypercharge to
$SU(2)_L \times U(1)_Y$), and $SU(3)_C$ as the color group
of quantum chromodynamics.\footnote{Again, in {\scmy Yang-Mills}
theories certain peculiarities arise, which will not be of our concern
here. Take for instance the {\scmy Higgs} mechanism in the electroweak theory.
It allows for non-zero masses of the gauge bosons by symmetry breaking
(since the gauge principle itself only leads to massless gauge bosons).
As yet, however, there is no experimental evidence for {\scmy Higgs} bosons.}
Let us briefly indicate the analogous application of the gauge principle
in the {\scmy Yang-Mills} case. Generally, in order to construct
an $SU(n)$ gauge theory, we use the GTL1
\be
\label{phase_N}
\Psi(x) \ \to \ \Psi'(x) = e^{i g_n \Lambda^a(x) \hat t^a} \Psi(x)
\equiv \hat U (x) \ \Psi(x)
\ee
of the fundamental spinor representation
$\Psi=(...,\psi^i,...)^T$, $i=1, 2 ... n$
analogous to (\ref{phasetrafo}).
The $n^2-1$ operators $\hat t^a$ generate the $SU(n)$ group;
each generator corresponds to a gauge potential $B^a_\mu$.
We obtain the non-abelian generalization of (\ref{pottrafo}), namely
\be
\label{pottrafo_N}
B^{\prime a}_\mu(x) \ \hat t^a
= \hat U(x) \ B^a_\mu(x) \ \hat t^a \ \hat U^+(x)
 - \frac{i}{g_n} \ \hat U(x) \ \partial_\mu \ \hat U^+(x) .
\ee
In this case the covariant derivative is given by
\be
\partial_{\mu} \ \to \ D_{\mu} = \partial_{\mu} + i g_n \ B^a_\mu \ \hat t^a ,
\ee
instead of (\ref{cov_der}) and the generalization of (\ref{fmunu}) yields
\be
\label{YMfieldstrength}
F^a_{\mu \nu} \ \hat t^a
= \Big( \partial_\mu B^a_\nu - \partial_\nu B^a_\mu
  - g_n f^{abc} B^b_\mu B^c_\nu \Big) \ \hat t^a
= - \frac{i}{g_n} \ [ D_\mu , D_\nu ] .
\ee
The bundle curvature $F^a_{\mu \nu}$ measures, so to speak,
the non-commutativity of the covariant derivatives.
In contrast to QED, the {\scmy Yang-Mills} Lagrangian
\be
{\cal L}_{YM} = - \frac{1}{4} F^a_{\mu\nu}(x) F^{a \ \mu\nu}(x) .
\ee
contains non-linear self-interaction terms of the kind
``$- g_n \partial B B^2$'' and ``$g^2_n B^4$'',
that is, the gauge bosons carry charge themselves.
The principal bundle structure is
$\setP( \Mink,SU(2)_L \times U(1)_Y )$ for the electroweak interaction,
and $\setP( \Mink,SU(3)_C)$ for chromodynamics.

\medskip

We may now ask for the genuine objects of quantum gauge field theories.
To be sure, the application of the gauge principle
shows the central role of the matter field $\psi$
as well as the gauge potentials $B^a_\mu$.
Without them the gauge argument is impossible,
since local gauge transformations GTL1 and GTL2 only
take effect on them. Moreover, in order to derive the
equations of motion and field equations, the corresponding
Lagrangians have to be varied with respect to $\psi$ and $B^a_\mu$.
Recall, however, that $\psi$ and $B^a_\mu$ are no gauge invariant
quantities and are, therefore, not directly observable!
By way of contrast, the gauge field strength $F^a_{\mu\nu}$
{\em is} gauge invariant -- as well as quantities bilinear
in $\psi$ such as the {\scmy Noether} current.
If we insist on the central role played by $\psi$ and $B^a_\mu$,
then this leads to the curious fact that, in gauge theories,
the genuine objects arising are themselves not directly observable.
Thus, it seems that we ourselves are faced with a dilemma -- similar
to {\scmy Field}'s one: the one horn being
that if we consider matter fields and gauge potentials
as genuine objects of gauge field theories (because they are
indispensible in deriving the theory from the Lagrangian or from
the gauge principle) we did not choose directly observable quantities,
but rather mathematical entities with no direct physical
significance.\footnote{One crucial remark, however: in {\scmy Yang-Mills}
theories the gauge potentials make a direct contribution to the
field strength tensor as can be seen from (\ref{YMfieldstrength}).
Moreover, in the {\scmy Yang-Mills} field equations
$$
(D_\mu F^{\mu\nu})^a(x)
= \partial_\mu F^{a \mu\nu}(x) - g_n f^{abc} \ B^b_\mu \ F^{c \mu\nu}(x)
= \jmath^{a \nu}(x) .
$$
the $B^a_\mu$ occur in direct combination with $F^a_{\mu\nu}$.
Therefore, in this case their physical significance is clearly
higher than for the $A_\mu$ potential in the abelian case.
}
However, if we restrict ourselves to directly observable
quantities only, we must give up the whole idea of gauge theories,
since per definition we can only apply the gauge principle
to non-gauge~invariant quantities. This is the other horn of the
dilemma.

We shall have to say more about this issue in the last section of
this paper. For the time being, we accept the first horn, that is,
we accept the occurence of objects in gauge theories
(seemingly even the primary ones!)
which, on the one hand side, are not directly observable,
on which, however, on the other hand side,
the whole idea of gauging is obviously based
and which are, insofar, truly indispensible.


\subsection{Gravitational gauge field theories}
\label{gravity}

The fundamental quantum field theories of the electroweak
and strong interactions clearly fit into the gauge theoretic framework
and possess a natural fiber bundle structure.
No such clarity can be found in the case of gravitational
theories.\footnote{By ``gravitational theory'' we henceforth
mean any theory which is based on the equivalence principle.
Besides general relativity the class of gravitational theories
includes unorthodox approaches of almost all imaginable
gauge groups of a four dimensional pseudo-{\scmy Euclid}ean spacetime
such as affine, linear and orthogonal groups,
as well as their corresponding covering and
supersymmetric groups and the diffeomorphism group
-- cf. Ne'eman (1980), Hehl et al. (1995).
The first gauge theory of gravitation was presented
by {\scmy R. Utiyama} (1956),
using the homogeneous {\scmy Lorentz} group $SO(1,3)$ as a structure group.
Later on, {\scmy F. W. Hehl} and others considered the full
{\scmy Poincar\'e} group $ISO(1,3)$ in a {\scmy Riemann-Cartan} spacetime
with curvature and torsion \cite{hehl_etal76,hehl_etal80}.
In most of the cases the experimental testability of alternative
theories of gravitation is beyond our current measuring accuracy
and, thus, the theoretical discussion is open for speculation.}
Nevertheless, we will argue that even gravitational theories
are best described in terms of gauge theories.
As a consequence, we will show that the underlying bundle
structure allows to distinguish naturally the physically
significant objects and constituents of these theories
from the merely mathematical ones.
Our position is by no means universal.
In fact, orthodox relativists are likely to claim
that a fiber bundle formulation of general relativity (GR)
is superfluous.\footnote{Compare, for instance, {\scmy J. Ehlers}:
`` ... {\em The formulation of the ``field kinematics'' of GR
in terms of principal bundles and their associated bundles
allows one to consider GR as a gauge theory ...
As far as I can see such gauge considerations have not led
to a deeper understanding of GR as such ...
I at least fail to see that the use of affine bundles with affine
(in {\scmy Cartan}'s sense) connections changes this fact, nor does it
help me to appreciate it more deeply. ... Of course these remarks are
not intended to pass any judgement on theories other than
{\scmy Einstein}'s, with or without gauge.}'' \cite{ehlers73}.}
This difference of opinion cannot be easily resolved.
However, we will try to demonstrate that a bundle formulation
of GR facilitates a better understanding of some outstanding
philosophical issues in gravitational theories.


\subsubsection{The gravitational gauge principle}
\label{gravgauge}

Let us, first, apply the gauge principle in the gravitational case.
For the sake of simplicity we restrict ourselves to GR,
which we shall present as a gauge theory of
the homogeneous {\scmy Lorentz} group $SO(1,3)$.
The generality of our conception will not be affected because
alternative gravitational gauge theories may be founded on
a gauge principle, too.\footnote{For different types of
gravitational gauge theories one basically has to plug in
different gauge groups. In fact, a more rigorous approach,
strictly following the gauge principle, leads to a gauge theory
of translations, since the gravitational field couples
to energy-matter, that is the corresponding {\scmy Noether} current
associated with global {\scmy Poincar\'e} translation covariance.}
We start with flat {\scmy Minkowski} space $\Mink$,
that is the interaction-free case in which no gravitation exists.
Hence, we consider the free geodesic equation,
\be
\label{free_geodesic}
\frac{d \theta^\mu(\tau)}{d \tau} = 0 ,
\ee
for a four vector $\theta^\mu(\tau) = \frac{d x^\mu(\tau)}{d \tau}$,
tangent to a timelike curve $x^\mu(\tau)$.
$\theta^\mu \equiv \theta_0^\mu$ together with a system
of three spacelike vectors $\theta_i^\mu$ represents a tetrad
$\theta^\mu_\alpha=(\theta^\mu_0, \theta^\mu_1, \theta^\mu_2, \theta^\mu_3)$.
Since we wish to derive gravity as a gauge theory of $SO(1,3)$,
the GTL1 analogous to (\ref{phasetrafo}) have the form
\be
\theta^\mu_\alpha(\tau)
= \hat L_\alpha^\beta(x) \ \theta^\mu_\beta(\tau)
= e^{ (\hat M^a)_\alpha^\beta \Lambda^a(x) }
\ \theta^\mu_\beta(\tau)
\ee
with $a=1...6$ generators $\hat M^a$ of $SO(1,3)$
and spacetime-dependent transformation parameters $\Lambda^a(x)$
(we set $x \equiv x^\mu(\tau)$).
In analogy to the {\scmy Yang-Mills} case (\ref{phase_N}), the Latin
index $a$ ``lives'' in the {\scmy Lie} algebra of the gauge group.
Note further that we must distinguish external space-time indices
$\mu, \nu \ldots$ from internal tetrad indices $\alpha, \beta \ldots$.
The former ``live'' in curved base space (holonomic indices),
whereas the latter ``live'' in local flat {\scmy Minkowski} space
(anholonomic indices), i.e. in the fibers of the tangent bundle
of the $SO(1,3)$ principal bundle over spacetime.
What is unique in the case of gravitation is the existence
of a natural mapping of the base space into the fiber space
(this mapping indices an isomporphism between
the associated vector bundle and
the tangent bundle of the spacetime base manifold).
The mapping allows us to distinguish between internal and external
indices. This phenomenon is sometimes referred to as the
{\em soldering} of base space and fibers.

Next, the gauge postulate, when applied to local {\scmy Lorentz} rotations
of the tetrads, leads to a covariant derivative $\nabla_\tau$,
\be
\frac{d}{d \tau} \theta^\mu_\alpha(\tau) \
\to \ \nabla_\tau \theta^\mu_\alpha(\tau)
= \frac{d}{d \tau} \theta^\mu_\alpha(\tau) + \Gamma^\beta_{\nu\alpha}
\frac{d x^\nu(\tau)}{d \tau} \theta_\beta^\mu(\tau) .
\ee
$\Gamma_\mu$ denotes the so-called {\em {\scmy Levi-Civita} connection}
with {\scmy Christoffel} symbols as components.
As in (\ref{pottrafo}), $\Gamma_\mu$ satisfies the GTL2
\be
\Gamma^\gamma_{\nu\alpha}(x)
= \hat L^\gamma_\alpha(x) \ \Gamma^\delta_{\nu\gamma}(x) \ (\hat L^{-1})_\delta^\gamma(x)
- \hat L^\delta_\alpha(x) \ \partial_\nu                 \ (\hat L^{-1})_\delta^\gamma(x) .
\ee
Finally, (\ref{free_geodesic}) becomes the geodesic equation
in curved spacetime
\be
\label{curv_geodesic}
\frac{d}{d \tau} \theta^\mu_\alpha(\tau) + \Gamma^\beta_{\nu\alpha}
\frac{d x^\nu(\tau)}{d \tau} \theta_\beta^\mu(\tau) = 0 .
\ee

Now we may compare GR with the quantum gauge theories
discussed above. Obviously, in the gravitational case, the gauge
potentials are given by the {\scmy Levi-Civita} connection $\Gamma_\mu$.
From this we may form the curvature tensor $R^\mu_{\alpha\beta\nu}$,
the so-called {\em {\scmy Riemann} tensor},
which represents the gravitational field strength.\footnote{Per
definition, the {\scmy Christoffel} symbols are symmetric in the
lower indices $\Gamma^\mu_{\alpha\beta} = \Gamma^\mu_{\beta\alpha}$.
Once we give up this restriction we may also have torsion
(besides curvature) as a gravitational field strength
in alternative gravitational gauge theories.}
A principal bundle with orthogonal structure group,
such as $SO(1,3)$, acting on tetradial frames, is called an
{\em orthonormal frame bundle} (or tetrad bundle, respectively).
Note that tetrads are represented by tangent vector fields,
i.e. sections in the associated tangent vector bundle.
Thus, in gravitational gauge theories, tetrads or reference frames
play a role analogous to the matter fields in quantum gauge theories.

Observe that, from the gauge theoretic point of view,
the metric tensor $g_{\mu\nu}$ is a derived object
which is built from the tetrads by means of
$g_{\mu\nu} = \eta_{\alpha\beta} \theta_\mu^\alpha \theta_\nu^\beta $,
with {\scmy Minkowski} metric
$\eta_{\alpha\beta} = diag(-1,1,1,1)$.
Observe further that the {\scmy Levi-Civita} connection is not
an independent object of the theory either (we shall represent it
in terms of derivatives of the tetrads). To say
that the connection is given in terms of the metric (which is given
in terms of the tetrads) is another way of expressing the soldering
of the bundle space (where the connection lives) and the base space
(where the metric lives).

We may associate with any reference frame an observer
in spacetime and, hence, a real physical system of ponderable
matter constituting measuring rods and clocks.
Note that observers in spacetime theories are, usually,
represented by mass points on timelike curves; cf. Earman (1974).
However, the gauge principle suggests that we represent observers
by tetradial frames, because the gravitational GTL1 act on tetrad indices.
Moreover, this construction is more general since it allows
for a representation of fermionic matter.\footnote{The close
connection between {\scmy Dirac} spinors and tetrads was already
used in {\scmy Weyl}'s 1929 paper.
From the modern aspect of encountering a quantum gauge
theory of gravitation, tetradial reference frames are also
important; cf. Rovelli (1991), Lyre (1998).}

By definition, the gauge principle allows us to derive
the gravitational equation of motion (\ref{curv_geodesic}).
In GR, the field equations are given by the
{\scmy Einstein} field equations which result from a Lagrangian
${\cal L}_{GR} = \frac{1}{2 \kappa} \sqrt{-g} R + {\cal L}_{Matter}$
linear in the field strength. To be sure,
a more general gravitational gauge approach shall
use a quadratic field Lagrangian; cf. Hehl et al. (1980).
The variation with respect to tetrads and connections as
genuine fields leads to the {\scmy Cartan} equations,
the true {\scmy Yang-Mills} equations of gravitation.
{\scmy Einstein}'s GR, therefore, only mimics the gauge theoretic
framework without really fitting in it.


\subsubsection{Objects and principles in general relativity}
\label{principles}

The key idea of the paper is to point out that
the fiber bundle formalism in gauge theories is fruitful
and that it clarifies the status of various theoretic objects.
We claim that even when gravitational gauge theories are concerned,
the bundle language provides a natural distinction
between the physically significant structures
and the merely mathematical ones.

Let us start with an important technical remark:
as we mentioned earlier, the structure group of
the fiber bundle is considered to be the gauge group $G$.
This terminology, though, can be seriously misleading
if we confuse the gauge group with the group $\cal G$ of local gauge
transformations. The latter one consists of spacetime-dependent
group elements and is, thus, infinite-dimensional.
In general, ${\cal G}$ is a subgroup of the automorphism group $\AutP$
of the gauge theory's principal bundle $\setP$.
The subgroup ${\cal G}_o$ of $\cal G$ of just the vertical automorphisms
is called the group of ``pure gauge transformations''.
Locally, $\cal G$ looks like a semidirect product of the
covariance group of the base manifold and the structure group.
Thus, we find ${\cal G} \cong ISO(1,3) \semisub SU(n)$
for {\scmy Yang-Mills} theories in {\scmy Minkowski} space.
In GR, however, ${\cal G}_o$ is already a subgroup
of $\DiffM$, the covariance group of the manifold,
i.e. we have ${\cal G} \cong \DiffM \semisub ISO(1,3) \cong \DiffM$.
{\scmy Andrzej Trautman} states this point as follows:
{\em ``... in the theory of gravitation, the group ${\cal G}_o$ of
`pure gauge' transformations reduces to the identity; all elements
of $\cal G$ correspond to diffeomorphisms of $\Man$.''}
\cite[p.~306]{trautman80a,trautman80b}.

This is the reason why it is difficult to distinguish between merely
mathematical and truly physical transformations and objects in GR.
From our bundle point of view, however, we are able to carry out
this distinction quite naturally.
We are faced with three different groups:
first, the covariance group of the spacetime manifold
($\DiffM$, in general).
Since it reflects a mere symmetry of the base space,
we may safely consider it as purely mathematical.
Second, the structure group $G$. It constitutes the fibers
and has significance insofar as it reflects the geometric
arena of the connections.
In \ref{gravgauge} we chose $G=SO(1,3)$, but as we already
mentioned, a more rigorous gravitational gauge approach uses the
{\scmy Poincar\'e} translation group. It can be shown that this group
is isomorphic to the group of local diffeomorphisms
(the reader may simply recall that a local diffeomorphism is nothing
but an infinitesimal point shift, i.e. a local translation).
Thus, in this case, we even get $G = \DiffM$.
Third, there is the group of local gauge transformations $\cal G$.
As we already emphasized, the GTL have physical significance
because of their role in the gauge theoretic framework:
they reflect the existence of an interaction field coupled
to the matter field and they have, thus, an active interpretation.
The curiosity of gravitational theories is that both the covariance
group as well as $\cal G$ are isomorphic to $\DiffM$
(and even $G = \DiffM$ for translational gauge theories).
The orthodox ``pure base manifold'' approach of GR, however, does
not allow to distinguish between $G$ and $\cal G$.\footnote{Concerning
the group of pure gauge transformations ${\cal G}_o$, however,
we must restrict our statement of the physical significance
of $\cal G$: pure gauge transformations, surely, do not have
a physical effect -- see our remarks in section \ref{resolve}.}
This distinction is one of the most important {\em conceptual}
advantages of the fiber bundle formalism in GR.
Similar ideas regarding gravitation have also been expressed
by {\scmy John Stachel} (1986).

The problem of the meaning of covariance within orthodox GR
is not a minor one.
The reader may recall that there is a longstanding question
of whether {\scmy Einstein}'s so-called principle of covariance
has any physical content\footnote{The prelude was given
by {\scmy Kretschmann}'s early objections 1917
-- see Norton (1993) for a general historical account.}
and, hence, whether a meaningful distinction can be made between
absolute and dynamical objects of GR.
One fruitful way of dealing with these questions has been
{\scmy James Anderson}'s proposal \cite{anderson67}.
Intuitively, an absolute object is one which remains invariant
or unchanged by the interactions of the theory under consideration.
However, it may itself change or influence other objects.
Examples are given by
the {\scmy Galilei} metric in {\scmy Newton}ian spacetime (NS) and
the {\scmy Minkowski} metric $\eta_{\mu\nu}$ in special relativity (SR).
The pseudo-{\scmy{Riemann}}ian metric $g_{\mu\nu}$ in GR,
on the other hand, represents a dynamical object.
To be sure, from our point of view the distinction between
absolute and dynamical objects is already captured in
the difference between base space and fibers:
dynamical objects ``live'' in the fibers,
whereas the absolute objects belong to the base manifold.

Now, {\scmy Anderson} as well as {\scmy Michael Friedman} proposed
to use the distinction between absolute and dynamical geometric
objects to characterize spacetime theories \cite{anderson67,friedman83};
see also Trautman (1973).
According to {\scmy Anderson-Friedman} any spacetime theory
is associated with a certain symmetry group, which is
{\em ``... defined to be the largest subgroup of the covariance group
of this theory, which is simultaneously the symmetry group of its
absolute objects. In particular, if the theory has no absolute object,
then the symmetry group of the physical system under consideration is
just the covariance group of this theory.''} \cite[p.~87]{anderson67}.
This is the case in GR. Any of the above mentioned examples of
spacetime theories can be characterized by a symmetry group
which preserves the absolute objects of the theory.
And conversely, any symmetry group accounts for a relativity principle,
which in the case of NS is the {\scmy Galilei} principle and
in the case of SR the special relativity principle.
For GR the {\scmy Anderson-Friedman}
approach has the advantage of showing that the principle of
general covariance has two different possible meanings:
Firstly, considered as a statement about the coordinate covariance
of spacetime theories, it is indeed physically vacuous, since any
spacetime theory which represents spacetime as a differentiable
manifold -- such as NS, SR, and GR -- may be written in a covariant
manner, i.e. form invariant under $\DiffM$.
Secondly, $\DiffM$ arises not only as a covariance,
but simultaneously as a symmetry group of GR.
This, however, is a highly non-trivial feature of GR;
it only obtains because GR does not contain absolute objects.
Therefore, this symmetry feature may be understood as the
{\em principle of general relativity}.
This principle has a clear physical significance,
unlike the principle of general covariance, which refers to
a merely mathematical (and not even characteristic) feature of GR.
It is the advantage of the {\scmy Anderson-Friedman} approach
that it points out the crucial conceptual difference
between these two principles.\footnote{Note, that the
{\scmy Anderson-Friedman} approach is designed to classify
spacetime theories in hierarchical order according to their
absolute objects (or their symmetry groups, respectively).
However, {\em ``... as {\scmy Robert Geroch} has observed,
since any two timelike, nowhere-vanishing vector fields defined
on a relativistic space-time are $d$-equivalent, it follows
that any such vector field counts as an absolute object...''}
\cite[footnote p.~59]{friedman83}.
Clearly, this raises a problem for the issue of classifying theories.
From our point of view, instead of using the distinction between
absolute and dynamical, we prefer to use the dichotomy between
the merely mathematical parts belonging to the base space
and the truly physical parts belonging to the fibers.
Indeed, one might even call these parts absolute and dynamical.
However, our distinction is not primarily intended to classify
spacetime theories, but rather to indicate their physical content.
With regard to timelike curves we clearly see that they do, indeed,
belong to the physical part of GR, since they are associated with
tangent vector fields, i.e. sections in the associated vector bundle.
In fact, reference frames, which may be associated with timelike vectors,
turn out to be the genuine objects of gravitational gauge theories
(see our remarks in section \ref{gravgauge}).
Thus, the bundle language shows that the only ``absolute'' entity
left in gravitational theories is the spacetime manifold itself
-- and its sole non-trivial property is that it admits a differentiable
structure (including global topology and the signature of its metric).}

From our bundle point of view, then, the confusion about the status
of the ``principle of covariance'' never arises.
As a mere base space property, general covariance has no physical
significance whatsoever.\footnote{Observe, though,
that general covariance is sometimes understood
to encode the universality of the gravitational coupling.
In the framework of bundle geometries this is more adaequately
expressed in the {\em soldering} of bundle and base space:
the gravitational interaction (bundle connection) represents
the geometry of the spacetime base manifold itself,
and, hence, every physical object {\em in} spacetime is affected.
Because of this feature, gravitation is distinguished from
other forces.}
On the other hand, to allow for a local gauge covariance in GR, that is,
to allow any observer to perform a transformation of local reference
frames with elements of $\cal G$, is only possible by introducing
a gravitational field to compensate for this local requirement.
This is the true physical content of the principle of GR which is,
thus, best understood as a gravitational gauge
principle.\footnote{However, as we already pointed out in footnote
\ref{gep}, the complete coupling structure of gauge theories
is based on the further assumption of an equivalence principle.
This is {\em a fortiori} true in gravitational gauge theories.}

To sum up: Gravitation can be understood as a gauge field theory proper.
Its genuine objects are tetradial reference frames, the bundle structure
is given by the principal bundle of orthonormal frames.
Bundles in gravitational theories, although they seem to be superfluous
due to the soldering of base space and fibers,
allow us to overcome conceptual problems concerning the meaning
of the various principles of relativity and covariance.
At the same time they help us to distinguish the truly physical
parts of these kind of theories.
In our opinion this proves that even in the case of gravitation
the fiber bundle formulation has clear advantages.


\newpage

\section{The resolution of ``Field's dilemma''}
\label{resolve}

In this last section we would like to conclude
with a focused attempt to answer various questions
which arise in connection with {\scmy Field}'s dilemma.

\subsection{Five questions}

\paragraph{Question 1:}
How does the fiber bundle approach affect a distinction between the
genuinely physical and the merely mathematical parts of physics?

\noindent
As we emphasized throughout the paper we believe that when gauge
theories are formulated in terms of fiber bundles
the merely mathematical part should ``live'' in the base space
while the truly physical part is contained in the fibers.
The base space is, essentially, a system of coordinates.
The need for such a ``coordinatization'' is a deep one -- it is
required whenever we wish to define a physical interaction.
Such a setting is mandated by the very concept of measurement
or observation: we can only observe and measure an interaction
if we can identify the location where it takes place.
The coordinate system, though, as useful as it is and as unavoidable
for the application of the theory, should not be considered,
in and of itself, a physical phenomenon or a collection
of physical objects.

 Another way of formulating the distinction between the mathematical
and the physical is the following. Given a mathematical space
which represents a gauge theory in its entirety
one should identify, first, the class of physical interactions
which appear in the theory.
The second step is to find out what the formal properties
of each interaction are, that is, those properties
which remain the same irrespective of location.
These formal features can usually be described as the operation
of a group which is called the structure group.
Finally, once the group is defined, we construct the equivalence
relation to which the group gives rise and describe the
total space as a principal bundle:
the space is ``quotientized'' with respect to the structure group.
The base space is then presented as a way of ``indexing'' the various
fibers in a smooth way, as {\scmy Sunny Auyang} has pointed out
\cite{auyang95}.
This indexing system does not need to have a physical interpretation
and it is perfectly consistent with a relational approach to the
spacetime base manifold. As we remarked earlier, the two points of view,
``coordinatization'' and ``quotientization'', are mathematically
equivalent with respect to the construction of fiber bundles.

As we have seen the vertical connections are what allows us to compare
the interaction in different regions. In and of themselves
they do not have any physical significance; they do, however,
allow us to see how does the interaction, which is schematically
represented by the operation of the structure group on the fibers,
propagate from one region to the next.

\paragraph{Question 2:}
In what sense is the merely mathematical part, which we identified with
the base manifold structures, dispensable?

\noindent
When we say that the base manifold is dispensable we do not mean that
we can do physics without a base manifold. We have had the occasion to
see that whenever we wish to discuss a local interaction we must use some
coordinate system. Even mathematically, the very idea of generalizing
the direct product necessitates some way of covering the space with
coordinate patches. There are, however, many different ways to coordinatize
a space. These different methods must be, in some sense, equivalent
(in mathematical terms the building blocks of one method must be
obtainable by a diffeomorphism from those of the other).
This, however, does not mean that the methods are identical and,
hence, we should not get rid of this ``layer of hidden structure''
too easily. First, there might be a way of distinguishing between
two diffeomorphic systems.
Second, as we have seen, there are cases where the diffeomorphisms
are ``active transformations'', that is, they have a significant
physical interpretation (this is the case, for example,
when gravity is construed as a gauge theory of local diffeomorphisms
which are mathematically isomorphic to local translations
-- see section \ref{gravity}).
Only in those cases where the diffeomorphisms do not have such
an interpretation (for example, when they represent
coordinate transformations) are we free to say that the
difference between diffeomorphic coordinate systems can be ignored.
It is precisely in this sense that a particular coordinate system is
dispensable. Any other diffeomorphic coordinate system would have served
just as well giving rise to a theory that is equivalent from a physical
point of view.

\paragraph{Question 3:}
Are the fiber bundle formulations of physical theories ``reasonably
attractive''?

\noindent
The question is whether our approach is aggressively revisionistic
and philosophically motivated or not.
It is a rather important question with wide implications.
To put it bluntly the question is whether physics itself admits
a natural distinction between the mathematical and the physical
or whether it takes a philosopher with a ``hidden agenda'' to accept
the formulations we offer because they permit such a distinction.

The first answer we would like to offer is that
the theories in the standard model (including gravitation)
admit a canonical fiber bundle formulation.
Moreover, the gauge approach seems to provide the only ``link''
between quantum field interactions and gravitation.\footnote{We
do not consider approaches beyond the standard model
such as supersymmetry, strings, etc.}
This aspect of unification is certainly one of the strongest arguments
for the use of fiber bundles in contemporary physics.
This point, though, does not settle the issue. There is a more
fundamental reservation, that is felt by many physicists, that the hyphaluted
mathematics of fiber bundles is superficial, that layers of unnecessary
mathematical structure are postulated, that fiber bundle formulations are,
simply, superfluous.

As we have seen before, one voice of dissent comes from the
orthodox general relativity theorists. This is not surprising.
As the reader undoubtedly knows the founding fathers of modern
gravitational theories, headed by {\scmy Einstein} and {\scmy Weyl},
attempted to ``geometrize'' physics.
This ambition precludes any rough and ready distinction between
the geometric (and hence, the merely mathematical) and the physical.
Indeed, as we remarked earlier even from a fiber bundle theoretic
point of view gravitation is a special case.
In the case of theories of gravitation the fibers are ``soldered''
to the base manifold.
This means that the extra degree of freedom which, in most cases,
allows for the introduction of the operation of a structure group
on the fibers does not exist in the case of gravitation.
Why, then, should we bother with the fiber bundle formulation?
Why not formulate gravitation in terms of manifolds with curvature
(and torsion) alone?
An even more fundamental worry is that the fiber bundles which
arise in gauge physics are, in most cases, mathematically trivial,
that is, they admit global sections and, therefore,
they are isomorphic to direct products.\footnote{\label{ab-note}
Let us mention, ever so briefly, that physical cases exist whose
formulation in fiber bundle theoretic terms is not trivial.
Some of them are known in the literature under the terms
geometrical or topological phases.
The best known case of this kind is the {\scmy Aharonov-Bohm} effect:
The quantum phase of a particle's wavefunction parallel transported
along a closed loop around a trapped magnetic field (such as a torroidal
magnet, for instance) shows a phase anholonomy which is experimentally
tested via interference patterns \cite{aharonov+bohm59}.
Anholonomy means that closed loops
on the base manifold are lifted to open curves in the bundle.
The presence of an anholonomy prevents one from defining
global cross sections. Therefore, a fiber bundle arising
in connection with the {\scmy Aharonov-Bohm} effect is non-trivial
(a more general account of physical anholonomies is given
by the theory of {\scmy Berry} phases; cf. Shapere and Wilczek, 1989).
Non-trivial bundles in quantum gauge theories may even also indicate
topological effects such as monopoles and instantons.
Most of them, however, are not vindicated experimentally
(at least from a foundational point of view).
This is the reason why we do not want to draw decisive arguments
from them. Moreover, as we mentioned already, we do not
consider approaches beyond the standard model to settle our point;
see, for example, {\scmy T. Y. Cao} on grand unified and
supergravity theories \cite[chap.~11.3]{cao97}.
Sure enough, the fundamental existence of such topological effects
would make our arguments concerning the significance of fiber
bundles even stronger, if not non-refutable!}
Does all this not mean that the fiber bundle formulation is superfluous?

We would like to offer a more general response to the
claim that the physics of fiber bundles is trivial.
As we have mentioned earlier there are three types of fiber bundles:
\bd
\item[(i) Trivial bundles with flat connections.]
In this case the connections are constant.
{\scmy Galilei}an spacetime, for example, can be viewed as a
fiber bundle of that type. It has absolute time $\setR$
-- the bundle's base space -- and relative space $\setR^3$
-- the fibers with structure group $O(3)$.
The bundle is globally isomorphic to the direct product
$\setR \times \setR^3$ and therefore trivial.
The flatness of the connections indicates space as a non-dynamical,
i.e. purely mathematical object.
Regarding type (i) we can safely say that the fiber bundle
formulation is completely superfluous.

\item[(ii) Trivial bundles with non-flat connections.]
In such cases the connections become dynamical quantities,
but the bundle space is still isomorphic to a direct product.
Both quantum gauge theories and gravitation give rise
to fiber bundles with non-flat connections.
We would like to suggest that the existence of non-flat connections
signifies the {\em physical non-triviality} (as opposed to their
triviality in the mathematical sense) of the fiber bundles involved.
Consequently, we believe that in those cases involving bundles
with non-flat connections the fiber bundle formulations are to be
taken seriously. This claim is, clearly, a contentious one;
for the time being we have no conclusive proof that it is defensible.
Can we make do with an argument showing that in all the known cases
the existence of non-flat connections has physical significance?
Certainly we cannot pretend to have exhausted the issue in these pages.

\item[(iii) Non-trivial bundles.]
Here, the bundle space is no longer isomorphic to a direct product.
In footnote \ref{ab-note} we used the opportunity
to briefly discuss these cases already.
There is no doubt that, in cases which fall under this category,
the bundle formulation is important, fruitful and, perhaps, indispensible.
But, as we mentioned in the above footnote, there may be some doubt
about the fundamental role these cases play in physics.
\ed

\paragraph{Question 4:}
Do all the mathematical objects which are defined on the fiber space
stand for measurable interactions or observable physical objects?

\noindent
Our aim in this paper was to show that we can assume that physical
phenomena have measurable consequences while, at the same time,
affecting a distinction between the merely mathematical and the truly
physical. Our argument was an inductive one.
We showed that our point of view is consistent with current physics.
Recall that the significant structures, from a physical point of view,
are matter fields, gauge potentials, gauge field strengths,
and local gauge transformations.
Therefore, what we need to show is that all these cases involve
only objects which live in the fibers.

We shall discuss them one after another.
\bd
\item[Matter fields.]
They are constructed as local sections in the vector bundles
which are associated with principal bundles and happen to be
the basic physical constituents of the gauge principle.
In the standard model the matter fields are given by the fundamental
elementary particle fields;
in gravitational gauge theories they are tangent vector fields
of the base manifold which are associated with reference frames.
The matter fields are not gauge invariant themselves;
only bilinear quantities as the {\scmy Noether} current
are directly observable. Nevertheless, there are
two reasons why we cannot do physics without them:
first, we may observe interference effects which cannot be explained
from the quantities bilinear in the matter field wave-functions.
Second, they are indispensible for explaining the gauge principle,
since local gauge transformations GTL1 act on them.
Therefore, matter fields and the vector spaces in which they live
(the fibers of the associated vector bundles)
have a certain physical significance.
From their conceptual role in the gauge theoretic framework we must
even consider them the building blocks of gauge theories.

\item[Gauge potentials.]
They are represented as coefficients of the connection forms
which split the tangent space of a principal bundle into a horizontal
and a vertical part. For physicists this concept is rather known
as ``covariant derivative''. They give rise to the gauge bosons.
Potentials are, again, not directly measurable,
but in the case of non-trivial bundles they will have
observational consequences in terms of topological effects.
They are, also, indispensible for an understanding of the gauge
principle, since local gauge transformations GTL2 act on them.

\item[Gauge field strengths.]
They are represented by the bundle curvature.
As gauge invariant quantities they are directly measurable as forces.
With respect to our three-fold distinction of bundle types above,
they give rise to type~(ii) bundles which we considered as sufficient
for explanatory purposes (since non-flat connections are
defined by non-vanishing curvature).

\item[Local gauge transformations.]
These transformations give rise to the structure group $G$
and, hence, to the group of local gauge transformations $\cal G$
(recall our remarks in section \ref{principles}).
The postulate of local gauge covariance,
as it is used in the gauge principle,
determines the form of the interaction coupling of the Lagrangian
(the GTL1 lead to the existence of an interaction gauge potential
with transformations GTL2; in other words, the GTL determine
the dynamics).\footnote{Again, as we remarked several times, the
true ``determination'' of the gauge potential as being physical
necessitates a further assumption, which may be formulated in terms
of a generalized equivalence principle (compare footnote \ref{gep}).}
In our opinion this constitutes a strong argument in favour
of an active interpretation of local gauge transformations.

On the other hand, the subgroup ${\cal G}_o$ of $\cal G$ of the
so-called pure gauge transformations has no physical significance,
since these transformations merely transform `vertically'
within an equivalence class of fiber points.
In other words, pure gauge transformations preserve the connections.
It is, of course, unavoidable to have an element of redundancy
within the bundle framework, since the whole idea of the
generalization of the direct product is rested on the existence of a
hidden layer of structure, namely the gauge group equivalence classes,
as we pointed out in section \ref{generalization}.
When we speak about active local gauge transformations,
we mean that they have a vertical as well as a horizontal component
(as encoded on the covariant derivative).
Note also, that our usage of ``active'' here essentially means
``physically effective'', i.e. changing the physical situation.
This, indeed, is the case when local gauge transformations are
concerned. Because of the gauge principle,
the postulate of local gauge covariance can be satisfied only
by introducing an interaction gauge field.
And this changes the physical situation.

\ed

In this respect we want to emphasize that the common view of
interpreting point transformations as active transformations
raises a different issue.
Those, who regard point transformations as active,
in the above-mentioned physical sense, are likely
to regard points and spaces as entities.
This touches on the issue of relationalism/substantivalism
which we will address in our last question.

\paragraph{Question 5:}
Are we committed to substantivalism with respect to bundle spaces?

\noindent
Substantivalism with respect to fiber bundle spaces is the doctrine
that bundle spaces, their constituents, bundle space points, and, hence,
also the base space itself, are genuine individuable entities.
This position raises many interesting philosophical questions.
One has to admit that, at a first glance, the fiber bundle formalism
can be utilized to justify this most extreme substantivalist position.
Our point, though, was a different one.
We wanted to affect a distinction between the truly physical parts
of fiber bundle gauge physics and those which are not;
and we identified the physical parts with the various fields and
transformations arising in the fibers
(as listed under the previous question).
This point, however, does not commit us to substantivalism.
In fact, the distinction we make between the mathematical and the
physical parts of gauge field theories does not commit us to any view
about the ontological status of mathematical entities (one may think
of the spacetime metric in GR as a physical object without taking it
to be a substance).
To be sure, questions concerning the status of fiber spaces
are of a novel kind and certainly very interesting.\footnote{It
is possible to develop a bundle space hole argument to rule out
bundle space substantivalism \cite{lyre99b} in analogy to
the well-known spacetime hole argument \cite{earman+norton87};
compare Earman (1989, chap.~8-9) also for comments on
the hole argument and {\scmy Field}'s substantivalism.}
Observe, though, that this issue is not logically related
to our approach. We are not committed to the view that fiber spaces
are themselves physically significant; we rather maintain that
the functions living therein, and not the spaces,
are genuine physical objects.

\subsection{Conclusion}

We hope to have demonstrated that our approach is one way of
resolving {\scmy Field}'s dilemma. Our method was an inductive one.
By looking carefully at a whole collection of gauge theories
we noticed that they all admit a natural distinction between
the physical and the mathematical parts.
This does not strike us as an accidental fact, although,
we have no way of proving it conclusively.
Therefore, on the basis of our observations, we conclude
that we obtained a general result: the physically
significant objects and quantities can be represented
as functions which are defined on the fibers
(which does not mean that, conversely, all of the structures
in the fibers are physically significant).
Our question of distinguishing between the mathematical and
the physical parts was originally motivated philosophically.
But it seems plausible to assume that the evolving degree
of unification in physics should entail a natural distinction
between these both parts in terms of the architecture
of our fundamental theories itself.
Fiber bundle gauge theories, we believe,
are a crucial step in this direction.


\newpage


\subsection*{Acknowledgements}

We would like to thank the people at the
Center for Philosophy of Science of the University of Pittsburgh
for their kind hospitality and support during our common stay
as Visiting Fellows during the 1999 spring term.
Special thanks to {\scmy John Earman} and {\scmy John Norton} (Pittsburgh)
as well as {\scmy Tim Oliver Eynck} (Amsterdam)
for their questions and corrections.
One of us (H.L.) was partially supported
by the Alexander~von~Humboldt-Foundation (Bonn).



\vspace*{10mm}

\bibliographystyle{apalike}


\end{document}